# Greeks were right : critical comments on Qbism .

**Author:** Marian Kupczynski

**Affiliation:**

Département de l'Informatique, Université du Québec en Outaouais (UQO), Case postale 1250, succursale Hull, Gatineau. Quebec, J8X 3X7 , Canada

Correspondence to: marian.kupczynski@uqo.ca

**Abstract :** In this short essay we reject the interpretation of quantum theory called quantum Bayesianism ( Qbism) which has been promoted recently by David Mermin in his essay published in Nature. According to Qbism quantum states are personal judgements of human agents. Physicists are not verifying their personal beliefs about their observations but search for the mathematical abstract description allowing to explain and to predict in a quantitative way the regularities observed (and those to be discovered) in physical phenomena which exist independently of the presence of any agent. We reject also the claim that Qbism explains properly the quantum nonlocality. The distant long range correlations can be correctly explained using a contextual statistical interpretation of quantum theory. We conclude that the Greeks were right to remove a perceiving subject out of Science.

In a recent essay published in Nature David Mermin[1] is promoting an interpretation of quantum theory called quantum Bayesianism ( Qbism) . We strongly disagree with this point of view and we give below arguments against it.

Qbism was proposed in 2002 by Carlton Caves, Christopher Fuchs and Rudiger Schack and became a subject of several papers and books.

In a paper: An Introduction to Qbism with an Application to the Locality of Quantum Mechanics '' published in 2013 the authors[3] claim that Qbism is <u>the</u> correct interpretation of quantum theory and that it solves all the famous paradoxes discussed during the last 80 years . Reading their article one may understand that Bohr, Schrodinger, Heisenberg and Einstein had not enough courage to accept a subjective view of probability . One may find also another claim in anonymous article in Wikipedia that John von Neumann was the first Qbist. All such claims are incorrect because fathers of quantum theory were convinced that quantum probabilities have nothing to do with subjective beliefs of agents.

The authors do not hide on the page 10 that they use a kind of circular reasoning : ''since the Qbist position… is an inevitable consequence of the subjective view of probability ( which is a correct one) … one could maintain that Qbism ( which the correct interpretation of QT) provides a validation of the personalist view of the probability''.

The subjective view of probability promoted by Bruno de Finetti may have some utility in different domains of science or in gambling but subjective beliefs of the agents or gambling bets have nothing to do with the probabilistic regularities observed in the physical phenomena.

In the paper[3] one finds several correct citations and statements of famous physicists but the correctness of these statements neither prove the correctness of reinterpretations of these statements given by the authors nor the correctness of Qbist and Cbist positions.

The aim of physics is to discover 'the laws'' of Nature governing objectively existing external world Physicists are not verifying their personal beliefs about their observations but search for the mathematical abstract description allowing to explain and to predict in a quantitative way the regularities observed (and those to be discovered) in physical phenomena which exist independently of the presence of any agent.

Of course we perceive the surrounding world by our senses and we probe it using experimental devices constructed by us. If our senses and brains were different we would probably grasp different aspects of the surrounding us world if we were able to do it.

It is amazing that the theories which we were able to create grasp not only the important laws of Nature but allow us to produce new materials, cure diseases and explore the Solar system. Perhaps our success is due to the fact that our senses and brains are the product of the evolution which had been governed by the laws of Nature which we were able only to discover in last centuries.

These laws of Nature governed physical phenomena long before Homo sapiens started their quest for the explanation of the observed phenomena. Stars were created and vanished, tectonic plates were moving, planets were orbiting around the Sun, seasons and tides were changing periodically. We can continue with examples from the biology: animal migration patterns were repeating, courtship rituals and mating were coded in genes; the evolution of species was driven by the environmental changes etc.

Classical physics and its mathematical models of the world are based on the assumption that our observations and measurements do not alter existing properties of the physical objects which we want to discover. For example a measurement of a position and a speed of a plane, by a radar method, does not affect its instantaneous position and speed. Similarly the true length of a table is not affected by the way we are measuring it. Any measurement, we make, is not precise but it was believed that by increasing the precision we could approach as close as we wanted the *true* value of a measured physical variable. This belief was proven to be incorrect when we started to study quantum phenomena, discovered incompatible physical observables and Heisenberg uncertainty relations.

In classical physics for a long time we were only studying properties of objects which we could perceive with our eyes and or with various magnifying glasses.

By studying different electromagnetic phenomena we discovered the existence of various fields and electromagnetic waves which we were unable to see. We discovered also the invisible world of atoms, molecules and elementary particles. We are only able to deduce the properties of this invisible world by studying its influence on the macroscopic phenomena observed using the macroscopic devices constructed for this purpose.

Following Planck and Einstein we discovered that the exchanges of the energy between the electromagnetic waves and the matter are quantized.

We discovered different amazing properties of this invisible world: the energy levels of atoms are quantized, the light behaves sometimes like wave and sometimes as a beam of 'photons', beams of the material particles (electrons, neutrons, molecules) passing by some crystals behave like waves .

By studying cosmic rays or by colliding beams of electrons, protons and ions we discovered hundreds of elementary particles and resonances. Using the Standard Model (SM)   particle physicists were able to explain quantitatively their properties and predict the existence of new particles which were discovered later.  Of course many questions in SM are not answered and comparison of the model with experimental data involves complicated Monte Carlo simulations containing many free parameters but it is the best model we have for now.

The strict localization of an elementary particle is impossible because an attempt to measure its position not only might  create several different particles but the particle whose position was measured could disappear.

In quantum phenomena invisible signals and particles prepared by some sources interact with the macroscopic devices and after substantial magnification produce clicks on various detectors.

Quantum theory provides an abstract mathematical model allowing: to make quantitative predictions about the energy levels of atoms and molecules,  to make probabilistic predictions concerning the statistical distribution of outcomes registered by various macroscopic devices etc.

The wave function is a mathematical entity not having any real existence and  it  is not an attribute of an individual physical system.  It is part of quantum mathematical model together with the operators representing various quantum observables.

Quantum probabilistic models contain often free parameters which can be treated as Bayesian priors. The best estimates of the unknown values of these parameters are obtained using the maximum likelihood model based on Bayesian approach.

However by no means quantum wave functions (state vectors) should be interpreted as the mathematical entity corresponding to the subjective belief of human agents.

Two high energy protons from cosmic rays, when they collide, are as likely to produce N mesons π   far away from human agents, as if they collided on the Earth, of course if the branching ratios for different reaction channels do not depend on the presence of gravitation.

Let us now comment  on Cbism[1]  promoted by David Mermin.

Inspired by Qbist position Mermin suggests that classical physics and a concept of a space- time should be also reinterpreted in terms of subjective experiences of agents. He also points out that a personal experience of a present moment (NOW) has no place in the special theory of relativity and that it is its drawback.

Our eyes allow us to perceive the surrounding us world in 3 dimensions and because we have an internal clock (heart beat), a personal perception of Now and a memory we discover that positions of objects in the surrounding us world change in time. It is well known that two observers watching the same scene or event may perceive it differently but there is something real going in the external world which creates these perceptions (or illusions).

A space –time is an idealized mathematical model abstracted from our subjective everyday observations which helped us to discover important laws of Nature , find regularities in the observed physical phenomena and actively change the environment in which we are living.

The space time is defined as a collection of point-like events (a position of a point-like particle at a given moment of time). In special relativity coordinates of these point-like events in different inertial frames are found using light beams and the radar method. It can be only done by well calibrated and synchronized devices installed by some human agents but once installed these devices can operate and register outcomes without a presence of any agent.

It is extremely useful idealization but as Mermin correctly pointed out there are no point –like events. The space –time loses its empirical foundations when we move from macro- to micro- world however several laws of Nature which we discovered in the macro-world are also valid in the micro-world.

Due to '' miraculous constancy of the speed of light in all reference frames'' the coordinates of the events in different reference frames are connected by Poincaré transformations (relativistic case) or by Galileo transformations ( non- relativistic case). The properties of space-time depend respectively on Poincare or Galileo group.

We discovered that that the invariance with respect to space translations corresponds to the conservation of the total linear momentum. The invariance with respect to time translation corresponds to the energy conservation. The invariance with respect to rotations corresponds to the conservation of the total angular momentum etc. We discovered the objective laws governing the external world.

These conservation laws are valid in the macroscopic world and they are valid in the *invisible world*. The formula $E=mc^2$ discovered by Einstein helped us to harness the energy liberated in nuclear reactions. The fast moving elementary unstable particles live longer than the same particles created by us in the laboratory what was also predicted by Einstein.

The validity of these laws does not depend whether some intelligent agents are there to verify them therefore the <u>Greeks were right to remove the perceiving subject out of Science</u>.

The discussion of quantum nonlocality in paper [3] contributes only to confusion. Statements such as: ''quantum mechanics is local because its entire purpose is to enable any single agent to organize her own degrees of belief about the contents of her own experience'' and ''quantum correlations are necessarily between time-like events'' are misleading and do not explain why strong correlations between the outcomes of distant random experiments [13-15] may and do exist.

A source is sending two signals to far away detectors and the clicks are produced and registered. After gathering several clicks the correlations between far away clicks can be estimated and these estimations outputted by on-line computers. The existence of outcomes does not depend whether some agents perceived these clicks or not.

The authors try to suggest that Qbism is the only theory able to elucidate various quantum paradoxes. It is not true: a contextual statistical interpretation [4-11] of QT, inspired by Einstein and Bohr writings, is free of all these paradoxes and is able to explain the long range correlations in spin polarisation experiments and the violation of Bell-type inequalities without invoking a mysterious quantum nonlocality[12-22].

Let us recall main assumptions of contextual statistical interpretation of QT :

1. A state vector or a density matrix represent an ensemble of identically prepared physical systems and are not attributes of individual physical systems .
2. Self-adjoint operators represent measured physical observables and together with density matrices allow to deduce probability distributions of the outcomes of repeated measurements of these observables .
3. No prediction is made on an outcome of a single measurement.
4. A state vector reduction is neither instantaneous nor non- local . A reduced state vector represents a different ensemble .
5. For EPR experiment a reduced state vector (for the system II) deduced from an entangled state of two physical system I+II describes only the sub-ensemble of the systems II being the partners of those systems I for which the measurement of some observable gave the same specific outcome.
6. Values of physical observables, such as spin projections of photons, are not predetermined attributes of physical systems but they are created during measurements in particular experimental contexts
   .

Qbists and Cbists forget that QT is much more than quantum information and that the physics is not interested in subjective experiences of various agents but it is trying to deduce the `laws of Nature' governing the external world .

Hans De Raedt , Mikhail I. Katsnelson and Kristel Michielsen introduced recently a mathematical framework of logical inference which can be used as a basis for establishing a bridge between objective knowledge gathered through experiments and their description in terms of (mathematical) concepts, thereby eliminating personal beliefs. The authors showed how quantum description emerges from this framework [23,24].

Epistemic interpretation of quantum mechanics and Qbism were also criticized ,from a different philosophical perspective, by Louis Marchildon [25,26] and by Michael Nauenberg[27]

**Conclusion:**

Our knowledge of the external world is only partial and based on the phenomena which we are able to create and/or observe by the devices which are available now. Therefore our knowledge of the external world (not a personal belief) is never absolute but relative[28] because it depends on our present tools of probing external world and it may change in time.

However this limited knowledge is much more than the subjective beliefs of some agents. This limited knowledge allowed us incredible technological achievements, contributed to the well-being of a human race and regrettably to the creation of the arms of mass destruction. We were so successful because <u>the Greeks were right to remove the perceiving subject out of Science.</u>

**Acknowledgements**: We want to thank Hans De Raedt and Louis Marchildon for reading a first draft of this paper and for valuable suggestions.